\documentclass[a4paper]{article}

\usepackage{icrc2013}

\title{ASTRI SST-2M Data Handling and Archiving System}

\shorttitle{ASTRI SST-2M Data Handling and Archiving System}

\authors{
L. A.~Antonelli$^{1,2}$,
S.~Lombardi$^{1,2}$,
F.~Lucarelli$^{1,2}$,
V.~Testa$^{1}$,
M.~Trifoglio$^{3}$,
D.~Bastieri$^{4}$,
A.~Bulgarelli$^{3}$,
M.~Capalbi$^{5}$,
A.~Carosi$^{1,2}$,
V.~Conforti$^{3}$,
A.~Di Paola$^{1}$,
S.~Gallozzi$^{1}$,
F.~Gianotti$^{3}$
M.~Perri$^{1,2}$,
G.~Tosti$^{6}$,
A.~Rubini$^{7}$,
S.~Vercellone$^{5}$
for the ASTRI Project.
}

\afiliations{
$^1$ INAF - Osservatorio Astronomico di Roma, Via Frascati 33, 00040 Monte Porzio Catone (RM), Italy\\
$^2$ ASI Science Data Center, Via del Politecnico, 00133 Roma, Italy \\
$^3$ INAF - Istituto di Astrofisica Spaziale e Fisica Cosmica di Bologna, Via P. Gobetti 101, 40129 Bologna, Italy\\
$^4$ Universit\'a di Padova, Dip. di Fisica e Astronomia, Via Marzolo 8, 35131 Padova, Italy \\
$^5$ INAF - Istituto di Astrofisica Spaziale e Fisica Cosmica di Palermo, Via U. La Malfa 153, 90146 Palermo, Italy\\
$^6$ Universit\'a di Perugia, Dip. Fisica, Via A. Pascoli, 06123 Perugia, Italy\\
$^7$ INAF - IAPS Roma, Via Fosso del Cavaliere 100, 00133, Roma, Italy \\

}

\email{angelo.antonelli@oa-roma.inaf.it, saverio.lombardi@oa-roma.inaf.it}

\abstract{
The ASTRI project is the INAF (Italian National Institute for Astrophysics) {\it ”flagship”} project developed in the context of the Cherenkov Telescope Array (CTA) international project. ASTRI is dedicated to the realization of the prototype of a Cherenkov small-size dual-mirror telescope (SST-2M) and then to the realization of a mini-array composed of a few of these units. The prototype and all the necessary hardware devices are foreseen to be installed at the Serra La Nave Observing Station (Catania, Italy) in 2014. The upcoming data flow will be properly reduced by dedicated (online and offline) analysis pipelines aimed at providing robust and reliable scientific results (signal detection, sky maps, spectra and light curves) from the ASTRI silicon photo-multipliers camera raw data. Furthermore, a flexible archiving system has being conceived for the storage of all the acquired ASTRI (scientific, calibration, housekeeping) data at different steps of the data reduction up to the final scientific products. In this contribution we present the data acquisition, the analysis pipeline and the archive architecture that will be in use for the ASTRI SST prototype. In addition, the generalization of the data management system to the case of a mini-array of ASTRI telescopes will be discussed.
}


\begin{document}
\maketitle

\section{Introduction}

ASTRI ({\it Astrofisica con Specchi a Tecnologia Replicante Italiana})~\cite{bib:pareschi} is a {\it flagship} project of INAF funded by the Italian Ministry of Education, University, and Research. This national project is strictly linked to the development of the international CTA project~\cite{bib:actis,bib:acharya}. INAF is currently developing an end-to-end prototype of the CTA small-size telescope in a dual-mirror configuration to be installed and tested at the INAF "M.C. Fracastoro" observing station in Serra La Nave (Mount Etna, Sicily)~\cite{bib:maccarone}. Data taking with this prototype is foreseen to start in 2014. The detailed description of the ASTRI project concerning the prototype structure, mirrors, camera, and scientific performance can be found in~\cite{bib:pareschi2,bib:catalano,bib:canestrari} and~\cite{bib:bigongiari}, respectively. A second step of the ASTRI project will be the deployment in 2016 of a mini-array, composed of a few SST-2M telescopes including the related infrastructures for data analysis and archiving, to be placed at the CTA Southern Site~\cite{bib:vercellone}. This mini-array will represent a remarkable improvement in terms of performance and could be the first seed of the future CTA project allowing to test most of the both hardware and software solutions to be adopted later by CTA. In the framework of the ASTRI project, the {\it Data Handling and Archiving System}, that is part of the Mini-Array System Software (MASS,~\cite{bib:tosti}) is responsible for both the on-site and off-site data processing and archiving of all the data produced by the different ASTRI sub-systems (Camera bulk data, calibrations, engineering data, housekeeping, auxiliary, etc.) as well as to provide the data access to the ASTRI community. Furthermore, the design of the ASTRI Data Handling and Archiving System is going to be compliant with the CTA Data Management (CTA-DM) and CTA Data Acquisition and Array Control (CTA-ACTL) requirements and guidelines. In this paper, we will describe the main characteristics of this system and of the software dedicated to it.

\section{ASTRI Data Handling \& Archiving System}

\begin{figure*}[!t]
  \centering
  \includegraphics[width=0.8\textwidth]{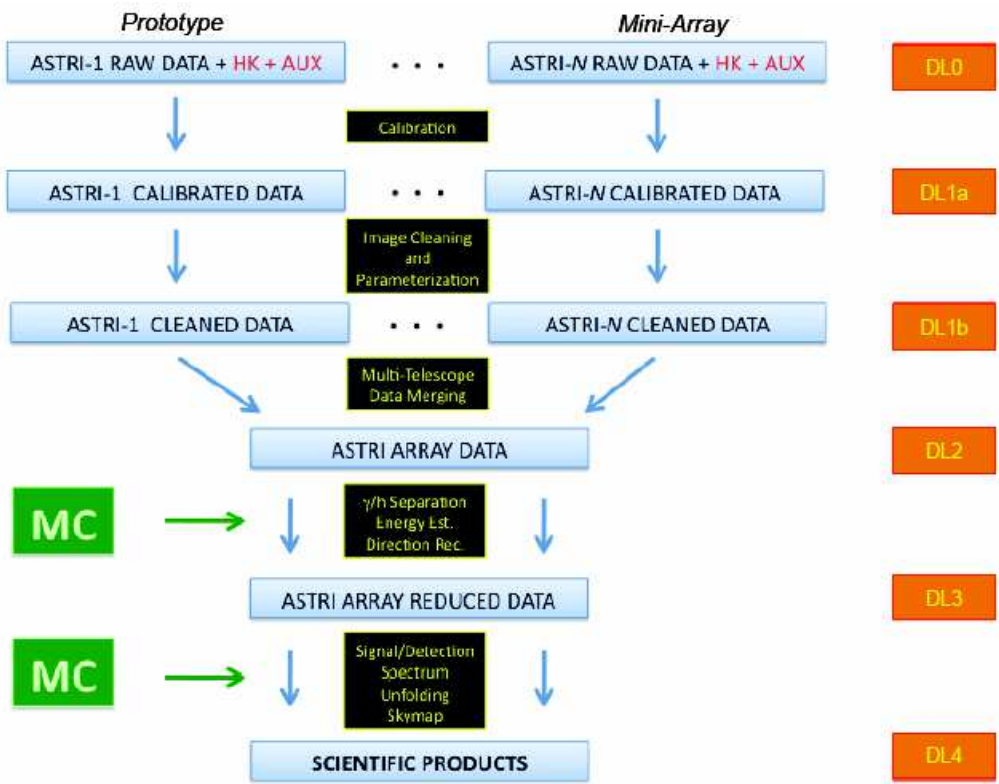}
  \caption{ASTRI reduction and analysis chain.}
  \label{dflow}
 \end{figure*}
 
The data processing for Cherenkov telescopes may generally be executed: {\it 1) in real-time on-site; 2) on-site at the end of each observation (on-site off-line); 3) in the data-centre at the end of the observation (off-site); 4) in the data-centre taking into account all data related to a specific target}. Each single processing can be considered in itself complete and it is composed by several steps starting from raw data and going on until the production of valid scientific information. Even if the logical structure of the data reconstruction and analysis process is the same for each of them the uncertainties are getting lower from mode 1 to mode 4 (due to the different calibration processes used). Moreover, the data processing activity is strictly connected with the Data Archive System and all the data obtained during the data acquisition, reduction and analysis process shall be properly stored and organized in the ASTRI project Data Archive. Following the different data processing activities itemized above, the archive will be composed by an On-site and an Off-site archive. The {\it ASTRI Data Handling \& Archiving System} is the system responsible for both the on-site and off-site data processing and archiving of all the data produced by the ASTRI project. It is initially conceived for the SST-2M prototype but looking forward to the SST-2M mini-array.

\subsection{ASTRI Data Flow}

The ASTRI Data reduction and analysis chain for both SST-2M prototype and mini-array is shown in Fig.~\ref{dflow}. The Data Levels are compliant with the definition of the CTA Data Levels~\cite{bib:ponz}. The main basic steps of the ASTRI telescope(s) analysis chain drawn in Fig.~\ref{dflow} are listed here below: 

\begin{itemize}
\item {\bf Raw Data (DL0).} The binary raw data, coming directly from the Data Acquisition (DAQ) system, are properly organized into FITS file.
\item {\bf Calibration (DL1a).} The signal for each pixel of the camera coming from the DAQ is properly calibrated and converted into number of photo-electrons (phe).
\item {\bf Image Cleaning and Parameterization (DL1b).} The aim of this analysis step is to remove, for each Cherenkov triggered event, those pixels containing light most likely not related to a Cherenkov event, and subsequently to perform the calculation of the Hillas parameters~\cite{bib:hillas} associated to the image of each event. 
\item {\bf Merging information coming from different telescopes (DL2).} Up to the image parameterization step, the analysis chain will run separately over the data streams of each telescope. At this level, N (where N is number of triggered telescopes for a specific event) different cleaned views of the same shower are available. Then, multi-telescope analysis will proceed in order to merge images belonging to the same Cherenkov event from the N data streams and to calculate further image parameters based on the multi-telescopes view of the same showers (stereo parameters).
\item {\bf Event selection ($\gamma$/hadron Separation, Energy Estimation, Direction Reconstruction) (DL3).} Once for each triggered event the image parameters (plus, in case, the stereo parameters) are available, classification algorithms will be used in order to estimate the nature of the events (hadron-like or $\gamma$-like) and to reconstruct their energy and the incoming direction. A gamma-candidate event list is built.
\item {\bf Scientific products (DL4)}. Scientific products such as sky maps, spectra, and light curves are finally obtained from the selected event list.
\end{itemize}

\subsection{ASTRI Data Acquisition Software}

\begin{figure}[ttt!]
\centering
  \includegraphics[width=0.5\textwidth]{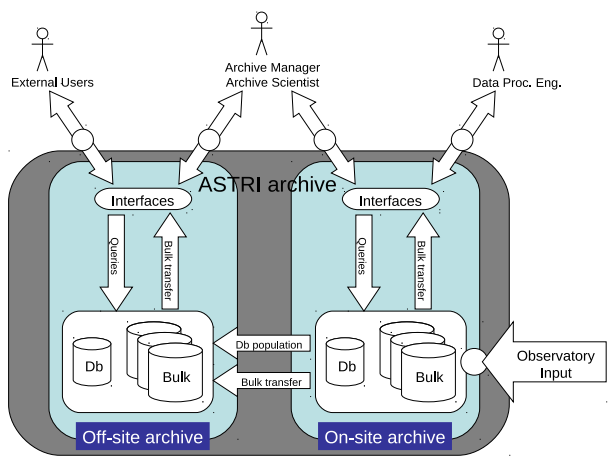}
  \caption{Logical structure of the ASTRI Data Archive.}
  \label{ada}
 \end{figure}

Each ASTRI camera is connected, through a point-to-point fiber Ethernet cable, to a dedicated computer, the ASTRI Camera Server, that runs the Detector DAQ Software (DDS). The camera server is in charge of collecting the bulk data sent by the Camera Back End Electronics (BEE) using the TCP/IP protocol. Each Camera bulk data stream consists of structured packets with various layouts pertaining to three main classes: notification packets, housekeeping packets, and science packets. A suitable packet header identifies the Camera data and allow distinguishing the various packet layouts. The size of each packet depends on the read-out mode. In science mode, the expected overall stream data rate for the typical event trigger rate of 600 kHz is of the order of 6-7~MB/sec. During ASTRI prototype nominal operations the DDS implements the data transfer from the BEE memory buffer over the DDS Event Builder to the on-site archive storing the raw data products. In the current design, both telemetry RAW data products and FITS raw data products are being evaluated. In addition, the DDS will support Commissioning and non-routine operations (e.g., troubleshooting) by providing local data storing, processing, and monitoring capabilities through a suitable operator Graphical Users Interface (GUI). In the mini-array configuration, according to the CTA design, the DDS will buffer the events coming from the ASTRI camera in the computer memory for several seconds and thus implement a secondary buffer, as required to interact with the Central Trigger and the Array Event Builder that will populate the RAW archive. Indeed, this buffering is required to perform on each forthcoming ``good event'' the following procedure: 1) to send the time tag to the Central Trigger and 2) to wait to know whether to send the read-out data to the Array Event Builder or to discard them. Software prototyping is being carried out to develop DDS modules embedded into ALMA Common Software (ACS,~\cite{bib:acs}) components so that DAQ controls, configuration and errors can be handled by the MASS at a higher control level. The Camera Server will include a component that treats the housekeeping data as individual ACS Basic Control Interface (BACI) properties and it will archive these data in the Telescope Monitor and Configuration Data Base (TMCDB). 

\subsection{ASTRI Data Reconstruction and Analysis Software}

The official software package for the ASTRI Data Reconstruction and Analysis is a collection of programs (classes and executables) written in C++ and using the standard HEASOFT CFITS−I/O libraries. It includes all the necessary algorithms to transform the raw data collected during a data-taking run, into information about the physics parameters of the observed primary incoming gamma-ray radiation. The data analysis chain is divided into several tasks, each of them performed by independent programs chained one with another by a pipeline. The data processing starts with the raw data for every triggered event, consisting of binary files containing the full information available per pixel (digitized signal amplitude versus time), plus camera housekeeping data. Other auxiliary information from the different subsystems (like e.g. the drive system(s), the (global) trigger system, the weather station, etc.) will be retrieved from the TMCDB. Throughout the analysis chain, the data are organized in files containing both an header with all the information relevant to the data reduction process and the {\it events}. Since no natural or artificial calibrators of VHE $\gamma$-rays are available for ground-based $\gamma$-ray detectors, the analysis chain must rely on Monte Carlo (MC) simulated $\gamma$-ray events to resemble $\gamma$-ray excesses in the ASTRI data. MC $\gamma$-ray events will be indeed essential for the $\gamma$/hadron separation (i.e. for the suppression of the overwhelming hadronic background in the data), the energy estimation of the events, and the calculation of the Effective Collection Area (a fundamental ingredient for the determination of the spectrum and light curve of a given observed source). The whole Data Reconstruction and Analysis process (as well as the MC production) is composed by a sequence of many operations, practically equal one to each other but repeated a large number of times. The time profiling of the code shows that most of the CPU time is devoted to the initial phases (calibration and cleaning) where the amount of data in input is vastly larger ($\times$100) than that of the following phases. The use of software specifically developed to take advantage of parallel computing architectures can instead reduce the calculation time for the data reduction, calibration, and final analysis by a factor of 10 to 100. In particular, the best approach, both in terms of initial investment and in terms of running costs, for parallel computing seem to be at present the GPUs (Graphics Processing Units). The ASTRI Data Reconstruction and Analysis software is being also developed using CUDA5 in order to exploit the GPU capabilities.

\subsection{ASTRI Data Archive}
The ASTRI SST-2M telescope(s) will produce data of different types: science data, calibration data, and housekeeping data. Other ancillary data will be also collected from the different subsystems providing information related to the observations (e.g. sky brightness, wind, humidity, etc.) and to the array configuration (for the ASTRI mini-array). MC simulations for the reduction and calibration of the acquired data will be also massively produced and archived. The principal components of the ASTRI Data Archive (ADA) are: 1) Data and Metadata; 2) software for the archive and the associated database(s) management; 3) hardware (i.e. storage devices, RAM and CPUs); 4) access data services. The ASTRI Data Archive will be formed by two separate units (Fig.~\ref{ada}):

\begin{itemize}
\item An {\bf on-site archive}, temporarily containing the raw data, the engineering (housekeeping and auxiliary) data, a fraction of MC data, and the products of the On-site analysis, including preliminary science products;
\item An {\bf off-site archive}, permanently containing the raw data, the Data Reduction products (up to high-level scientific products), and the definitive engineering archive. A dedicated branch will collect all the MC simulations produced for ASTRI.
\end{itemize}

 \begin{figure}[ttt!]
  \centering
  \includegraphics[width=0.5\textwidth]{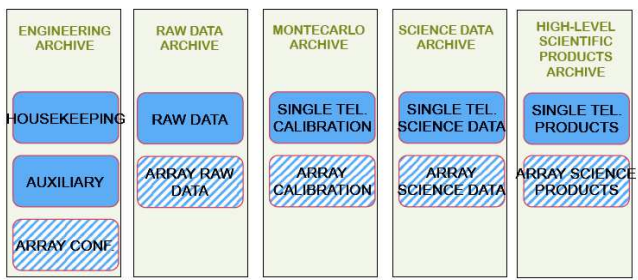}
  \caption{ASTRI Data Archive for both SST-2M prototype and mini-array.}
  \label{ada2} \end{figure}

These units are composed by five main logical levels compliant to the CTA observatory requirements (Fig.~\ref{ada2}):

\begin{enumerate}
\item {\bf Raw Data archive.} It contains the definitive archive of the raw data produced by the arrays and
all data products generated from the reduction pipeline up to DL2.
\item {\bf Science archive(s).} It includes reduced DL3 data and DL4 products (sky maps, light curves etc.) as
well as real-time analysis results.
\item {\bf Engineering archive.} It contains the archive of calibration data as well as the housekeeping and
auxiliary information.
\item {\bf Monte Carlo archive.} It contains all the MC data simulated for the different array
configurations.
\item {\bf Final Products archive.} High-level scientific results to be compared to other products from other
observatories at other wavelengths.
\end{enumerate}

The ASTRI prototype is expected to produce an amount of data of $\sim$ 0.6\,TB/night, while the mini-array shall produce around 3\,TB/night. Nevertheless, for the CTA Observatory, assuming a 15\% observation time over the whole year, a maximum volume of data per day is expected to be $6(\mbox{North-CTA})\mbox{TB}+17(\mbox{South-CTA})\mbox{TB} \simeq 24\mbox{\mbox{TB}}$ per day corresponding to a maximum volume of data per year of $0.7(\mbox{North-CTA})\mbox{PB}+1.9(\mbox{South-CTA})\mbox{PB} \simeq 3\mbox{\mbox{PB}}$ per year. In such a scenario a customized solution in order to easily manage such a huge archive is highly desirable. The development of a distributed archive system could be a suitable solution to CTA and the mini-array Science Archive could represents a unique opportunity to test it on an array configuration similar to the CTA observatory configuration. 

\section{ASTRI Engineering Archive}
The engineering archive will be organized under a parallel Data Base engine using the ACS TMCDB. It will store both the static configurations and the monitoring data for all the subsystems that will be made available for the data analysis. At the present time, the fine details (e.g. what DB engine to use) are still under discussion within the ASTRI community and among CTA partners and ESO in view of the adoption of a standard solution to deal with all the subsystems monitoring data.

\section{Conclusions}

The ASTRI SST-2M telescope is an end-to-end prototype of Small Size Telescope also including the development of both the Scientific Data Analysis Software as well as the Scientific Archive. The solutions adopted for both the ASTRI Data Analysis Software and Archive are compliant with the requirements of the Cherenkov Telescope Array international project. At the same time, also new possible solutions are investigated such as the use of GPU in the data reduction chain and a distributed archive for data archiving. The further evolution of the ASTRI project from the prototype to a mini-array will represent the unique opportunity to test these solutions on a configuration very similar to the one that will be realized on CTA observatory.

\vspace*{0.5cm}
\footnotesize{{\bf Acknowledgment:}{This work was partially supported by the ASTRI Flagship project financed
by the Italian Ministry of Education, University, and Research (MIUR) and led by the Italian National Institute of Astrophysics (INAF).
We also acknowledge partial support by the MIUR Bando PRIN 2009. LAA, AC and, FL are also supported by ASI through the 
INAF-ASI agreement for the activities of ASDC.}}

\end{document}